\documentclass[%
aip,
jmp,%
amsmath,amssymb,
preprint,%
]{revtex4-1}

\usepackage{graphicx}
\usepackage{dcolumn}
\usepackage{bm}
\usepackage{multirow}
\usepackage{color}
\usepackage{xfrac}

\begin{document}

\preprint{AIP/123-QED}

\title[]{Controlling the Frequency-Temperature Sensitivity of a Cryogenic Sapphire Maser Frequency Standard by Manipulating Fe$^{3+}$ Spins in the Sapphire Lattice }

\author{K. Benmessai}
\affiliation{ARC Centre of Excellence for Engineered Quantum Systems, School of Physics, University of Western
Australia, 35 Stirling Hwy., Crawley 6009, Western Australia
}
\email{karim.benmessai@uwa.edu.au}
\author{D.L. Creedon}
\affiliation{ARC Centre of Excellence for Engineered Quantum Systems, School of Physics, University of Western
Australia, 35 Stirling Hwy., Crawley 6009, Western Australia
}
\author{M. Mrad}
\affiliation{FEMTO-ST Institute, Time and Frequency Dept., 26 Rue de l'\' Epitaphe, 25 030 Besan\c con Cedex}
\author{J.-M. Le Floch}
\affiliation{ARC Centre of Excellence for Engineered Quantum Systems, School of Physics, University of Western
Australia, 35 Stirling Hwy., Crawley 6009, Western Australia
}
\author{P.-Y. Bourgeois}
\affiliation{FEMTO-ST Institute, Time and Frequency Dept., 26 Rue de l'\' Epitaphe, 25 030 Besan\c con Cedex}
\author{Y. Kersal\' e}
\affiliation{FEMTO-ST Institute, Time and Frequency Dept., 26 Rue de l'\' Epitaphe, 25 030 Besan\c con Cedex}
\author{V. Giordano}
\affiliation{FEMTO-ST Institute, Time and Frequency Dept., 26 Rue de l'\' Epitaphe, 25 030 Besan\c con Cedex}
\author{M.E. Tobar}
\affiliation{ARC Centre of Excellence for Engineered Quantum Systems, School of Physics, University of Western
Australia, 35 Stirling Hwy., Crawley 6009, Western Australia
}

\date{\today}

\begin{abstract}
To create a stable signal from a cryogenic sapphire maser frequency standard, the frequency-temperature dependence of the supporting Whispering Gallery mode must be annulled. We report the ability to control this dependence by manipulating the paramagnetic susceptibility of Fe$^{3+}$ ions in the sapphire lattice. We show that the maser signal depends on other Whispering Gallery modes tuned to the pump signal near 31 GHz, and the annulment point can be controlled to exist between 5 to 10 K depending on the Fe$^{3+}$ ion concentration and the frequency of the pump. This level of control has not been achieved previously, and will allow improvements in the stability of such devices.
\end{abstract}



\maketitle


\section{Introduction}

The Whispering Gallery Mode Sapphire Maser is a stable microwave oscillator, which makes use of paramagnetic Fe$^{3+}$ ions 
in ultra-low loss HEMEX cylindrical sapphire crystal at low temperature, and has been described in detail previously  
\cite{Pyb2005APL, Pyb2006IJMP, Benmessai2005EL, Benmessai2008PRL, Creedon2009IEEE}. 
The fractional frequency instability has been demonstrated to be as low as $\sigma_y 
(1s<\tau<100s) = 10^{-14}$. This instability is normally measured at the frequency-temperature 
turnover point (or annulment temperature) of a high Q-factor $(>10^9)$ Whispering Gallery Mode (WGM) excited in the sapphire, 
where the effects of temperature fluctuations on frequency are 
nullified to first order by the magnetic susceptibility of residual paramagnetic impurities\cite{Kovacich1997JPDAP, Giles1991JPDAP}. 
Many publications describe the application of this self-compensation technique 
\cite{Kovacich1997JPDAP, Giles1991JPDAP, Hartenett2007PRB, Dick1995IEEE, Dick1998IFCS, Wang2002IEEE, Naimi2005, tobar1999},
and all show that the annulment temperature is dependent on the presence and relative 
concentrations of paramagnetic ions such as Cr$^{3+}$ with an Electron Spin Resonance (ESR) 
at 11.4 GHz, Mo$^{3+}$ (100 GHz) and Ti$^{3+}$ (1 THz). The ions are substitutionally included 
in the crystal lattice during the crystal growing process and occur 
unintentionally. 

In this work we show that the frequency-temperature annulment for the Maser depends predominately on the Fe$^{3+}$ ions  and can be controlled by manipulating the number of ions involved in the maser process. We compare the behavior of two crystals - C1 (concentration of active Fe$^{3+}$ ions = 10 ppb) and C2 (100 ppb). We also report the observation of an upper limit in temperature of 30 K for operation of the maser. This limit is explained using a Boltzmann distribution for the active ion, and we show that the populations of the energy levels are so close at this temperature that the effect of the pump at 31 GHz used to excite the maser at 12 GHz is negligible. 

Controlling spins in such high-Q cavities could also have other potential applications. For example, it has recently been suggested that HEMEX sapphire with paramagnetic impurities could have applications for quantum measurement at millikelvin temperatures\cite{Creedon2010PRB, Creedon2011APL}, such as qubits or quantum memory applications\cite{Tim,PhysRevLett.92.076401,PhysRevLett.105.140502,PhysRevLett.105.140501,FuchsNatPhys,PhysRevLett.105.140503,PhysRevB.82.024413}. 

\section{Maser Description}\label{Section:Maser}
\begin{figure}[h!]
\centering
\includegraphics[width=0.5\textwidth]{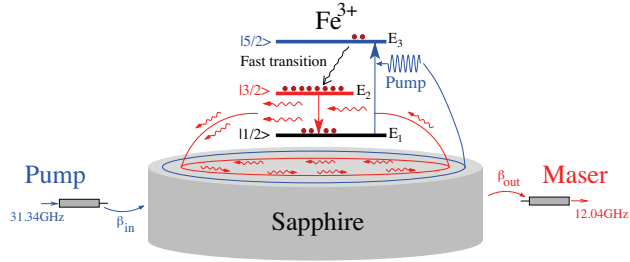}
\caption{\label{fig:1} \it{\small{Principle of operation: We apply a signal at 31.34 GHz, coinciding with a WG
mode,
to pump the ions from the lowest energy level $\left|1/2\right\rangle$ to the third $\left|5/2\right\rangle$. The ions then 
(non-radiatively) relax to the second energy level $\left|3/2\right\rangle$. A population inversion is obtained 
between the two lowest energy levels, and stimulated emission can be achieved at 12.04 GHz, 
which is enhanced by another WG mode coincident in frequency.}}}
\end{figure}

The maser scheme is based on the three energy levels of Fe$^{3+}$ ions at zero applied 
DC magnetic field in the sapphire lattice of a WGM resonator (Fig.~\ref{fig:1}). The concentration of active ions used
to 
create the effect is about 10-100 ppb, whereas the total concentration is on the order of 1 ppm. 
The bandwidth of the ESR is $\Delta \nu_{\small{\textrm{Fe}^{3+}}}\approx 27$ MHz, where the WG 
mode width is only $\Delta \nu_{\tiny{\textrm{WG}}}\approx 10$ Hz at 4.2 K. \\

The maser signal frequency $\nu_{\small{\textrm{op}}}$ is fixed at the frequency
$\nu_{\tiny{\textrm{WG}}}$ of the WG mode
involved in the process. It is well known that the frequency $\nu_{\tiny{\textrm{WG}}}$ is strongly
sensitive
to temperature variations and needs to be precisely controlled. In this kind of resonator, 
$\nu_{\tiny{\textrm{WG}}}$ possesses a frequency-temperature turnover point (also referred as the annulment or 
annulment temperature) above 4.2 K, although annulment temperatures as low as 90 mK have been recently measured \cite{Creedon2011APL}. 
The frequency variations can be described by \cite{Kovacich1997JPDAP}:
\begin{equation}\label{eqn:1}
\frac{\nu_{\tiny{\textrm{WG}}}-\nu_{0}}{\nu_{0}}=AT^4+C(T,\nu)
\end{equation}
where $\nu_0$ is the mode frequency at 0 K without ions, and $A$T$^4$ is the thermal 
sensitivity due to the sapphire dilation ($A\approx10^{-12}$ K$^{-4}$ for our modes). The other 
term $C(T, \nu)$ is thermal sensitivity due to the paramagnetic ions and is defined by:
\begin{equation}\label{eqn:2}
C(T, \nu)=\frac{1}{2}\sum_i \eta_i (\nu) \chi_{0}^{(i)}\frac{(2\pi\tau_2^{(i)})^2
\nu_i(\nu_i-\nu)}{1+(2\pi\tau_2^{(i)})^2 (\nu_i-\nu)^2}
\end{equation}
This term depends in reality on both temperature and operational frequency. The annulment temperature 
is then calculated by considering the maxima of Eqn. \ref{eqn:2}, where the index $i$ labels the different paramagnetic species present in the lattice.
The magnetic filling factor $\eta_i$ of the mode describes its magnetic field distribution 
in relation to the $i^{\text{th}}$ species of paramagnetic ion, and $\chi_0^{(i)}$ is the DC magnetic susceptibility of 
that ion. As the operating frequency is fixed at around 12.04 GHz, only the effect of the Fe$^{3+}$ is considered (i.e. the summation in Eq. \ref{eqn:2} is only performed over a single ion $i=\text{Fe}^{3+}$), and the annulment temperature ($T_{inv}$) can then be expressed as follows\cite{Kovacich1997JPDAP, mathese}:
\begin{equation}\label{eqn:3}
T_{inv}=\left( \dfrac{(g \mu_B)^2 \mu_0 \pi\tau_2 \nu_{12}\sigma_{12}^2 N_{int}}{48 k_B A} \right
)^{\dfrac{1}{5}}
\end{equation}

where the Bohr
magneton $\mu_B=9.274 \times 10^{-24}$ J.T$^{-1}$, the Boltzmann constant $k_B=1.38\times10^{-23}$ J.K$^{-1}$, 
the Land\' e g-factor $g\approx 2$, $\sigma_{12}^2\approx 2$ is a
characteristic constant of the absorption rate of the ions at 12.04 GHz (calculated from the Fe$^{3+}$ Sapphire spin
Hamiltonian parameters)\cite{Bogle1958, Siegman, mathese}, the spin-spin relaxation time $\tau_2$ is related to the 
width at half resonance of the ESR ($\Delta\nu_ {12}=\sfrac{1}{\pi \tau_2}$). 
If two spins are in phase, and one absorbs a photon causing a phase shift, $\tau_2$ is the necessary time for the spins to 
be in phase again.
\begin{figure}[h!!!!!!]
\centering
\includegraphics[width=0.5\textwidth]{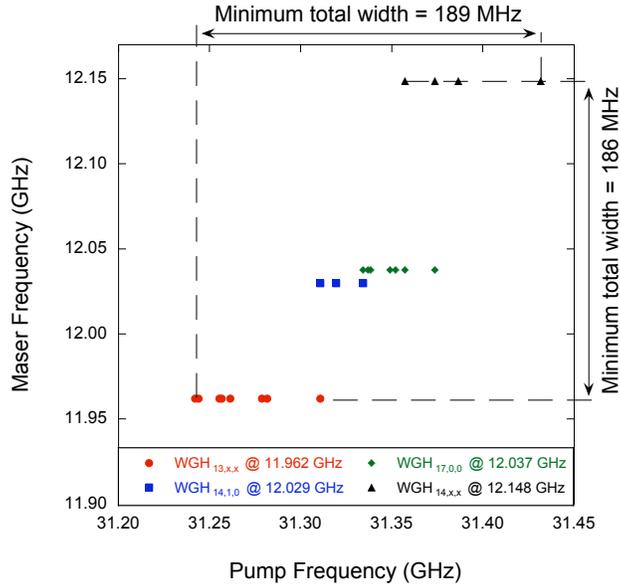}
\caption{\label{fig:2}\it{\small{The active maser bandwidth over which we can generate a signal is of order 180-190 MHz for crystal C2. Masing occurs at discrete 
frequencies where WGMs exist between 11.95 to 12.15 GHz, which are in turn pumped at discrete frequencies between 31.24 to 31.44 GHz 
where higher frequency WGM's exist.}}}
\end{figure} 

Fig.~\ref{fig:2} shows the bandwidth for active maser operation in crystal C2, which is more than 186 MHz for the 12 GHz and 
31 GHz ESR levels. We can note from Fig.~\ref{fig:2} that the maser can operate in a multimode configuration. For example, when the pump frequency is in the range 31.350 GHz - 31.375 GHz, two 
signals can operate at once at 12.037 GHz and 12.149 GHz. In the rest of the paper, the characterization of the turnover 
temperature has been investigated for single-mode maser operation. We consider then in this paper $\Delta\nu_{12}=27$ MHz, the width measured by Bogle and Symmons \cite{Symmons1962}.
$N_{int}$ is the concentration of the active ions involved in the annulment temperature, in units of ion/m$^3$. 
However, it is convenient to express it as a fraction 
of the Al$_2$O$_3$ in the crystal, i. e. $2.35\times 10^{22}$ ion/m$^3\equiv\ $1 ppm. This concentration is 
related to the total ion concentration in the lattice as follow:
\begin{equation}\label{eqn:}
N_{int} = \zeta \times N_{total}
\end{equation}
where $\zeta$ is the fraction of the ion concentration involved in the proces ($0\le \zeta \le 1$) and $N_{total}$ the total ion concentration 
in the lattice. When $\zeta=0$, no frequency-temperature annulment is observed, and when $\zeta=1$ all the ions in the lattice participate in the 
annulment process.
\begin{figure}[h!]
\centering
\includegraphics[width=0.45\textwidth]{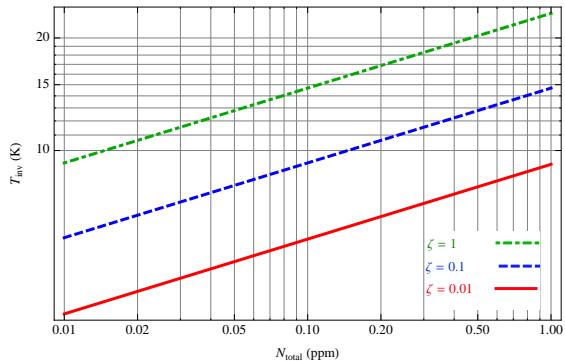}
\caption{\label{fig:3} \it{\small{Evolution of the annulment temperature with the total Fe$^{3+}$ concentration for
different values of the fraction $\zeta$.}}}
\end{figure}
Fig.~\ref{fig:3} shows that the annulment temperature $T_{inv}$ increases with the total number of ions $N_{total}$.


\section{Measurements}
\begin{figure}[h!]
\centering
\includegraphics[width=0.5\textwidth]{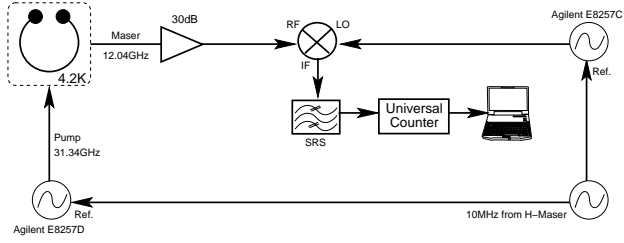}
\caption{\label{fig:4} \it{\small{Experimental setup: The crystal is a cylindrical HEMEX sapphire, 30mm 
height and 50mm diameter, placed in a cylindrical cavity and cooled to 4.2 K using a cryocooler.}}}
\end{figure}

To characterize the maser signal, we amplify it by 30 dB and then compare it to a signal from 
a synthesizer referenced to a hydrogen maser. The beat note is then amplified and band pass 
filtered before being counted (Fig.~\ref{fig:4}). The accuracy of the temperature control at the 
resonator cavity is about 2 mK. 


\subsection{Behavior of C1 with low concentration of Fe$^{3+}$}
\begin{figure}[h!]
\centering
\includegraphics[width=0.45\textwidth]{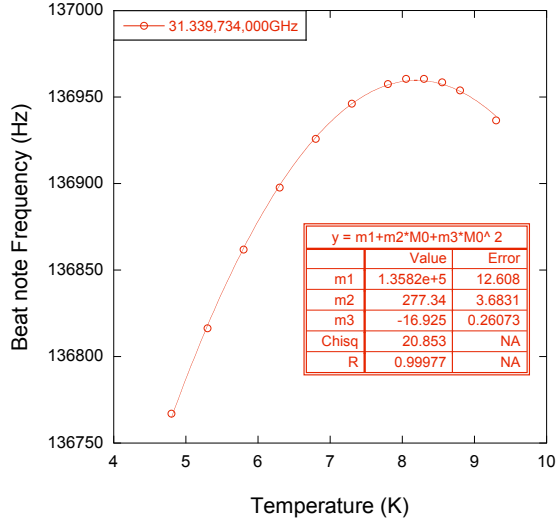}
\caption{\label{fig:5} \it{\small{Evolution of the maser frequency (for C1) at 12.038 GHz with 
temperature for a fixed pump at 31.339,734 GHz. From the fit, the annulment temperature is 8.193 K.}}}
\end{figure}

The maser characterized in this section oscillates at the WGH$_{17,0,0}$ mode frequency (12.038 GHz) of crystal C1, and can be excited using a pump signal corresponding to any of several WG modes 
around 31.34 GHz. The frequency of the WG mode that enhances the maser signal shows a turnover temperature $T_{inv}$ around 7.8 K, with no 
dependence on the input power applied to the crystal. The evolution of the maser frequency with temperature is 
shown in Fig.~\ref{fig:5}.

\begin{figure}[h!]
\centering
\includegraphics[width=0.45\textwidth]{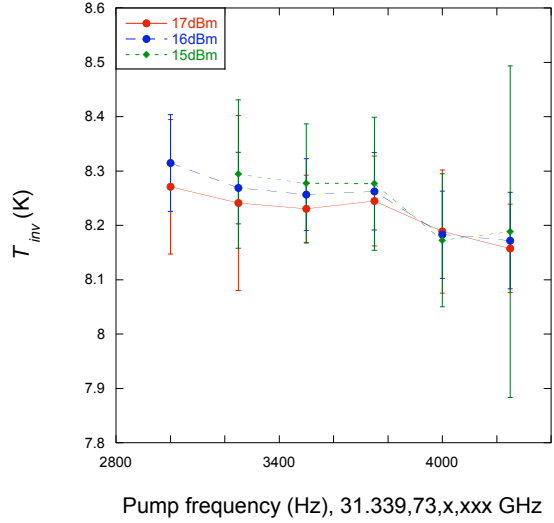}
\caption{\label{fig:pin2} \it{\small{Evolution of the annulment temperature for different values of the pump power and frequency. The pump mode chosen was at 31.339 GHz.}}}
\end{figure}

The excitation frequency of each pump WG mode was varied, and the evolution of the maser frequency with temperature was observed at various levels of pump power. The frequency was changed in steps of 250 Hz at various levels of pump power. As Fig.~\ref{fig:pin2} shows, $T_{inv}$ is independent of the pump power and the pump frequency for each pump mode. For example, the 31.339  GHz pump mode has an annulment temperature which stays constant at 8.25 K. 

\begin{figure}[h!]
\centering
\includegraphics[width=0.45\textwidth]{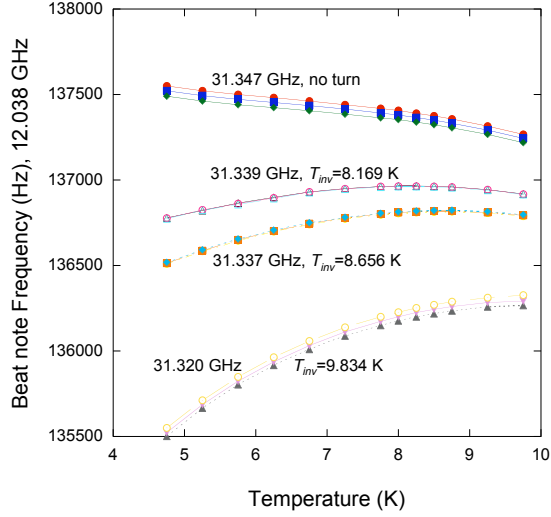}
\caption{\label{fig:6} \it{\small{Evolution of the maser signal frequency with temperature for different 
pumps at 31.3xx GHz. For each pump $T_{inv}$ is independent from its frequency and the power.}}}
\end{figure}

Despite the fact that $T_{inv}$ is independent of the input power and frequency of the pump 
signal (for a particular choice of pump mode), its value does change when a different 
pump WG mode is selected, as shown in Fig.~\ref{fig:6}. The $T_{inv}$ point is measured  for 
the pump frequency range 31.320 - 31.339 GHz. No turnover 
point is observed at higher frequencies, due to the annulment temperature residing outside the 
measurable temperature range (below 4 K). $T_{inv}$ can be calculated for each pump by a simple 
second order polynomial equation, as shown in Fig.~\ref{fig:5}.

For each pump at 31 GHz, the population of active ions $N_{maser}$ used to create the maser signal is different from that
fraction of ions that determine the $T_{inv}$ for the mode. The active ion concentration was deduced for each $T_{inv}$ 
from the maser output power calculated by Benmessai et al. \cite{Karim2010}, as shown by Eq. \ref{eqn:4},
\begin{widetext}
\begin{equation}\label{eqn:4}
N_{maser}=\dfrac{1+\beta_{out}}{\beta_{out}}\left[ \dfrac{\beta_{out}}{Q_0} \dfrac{3 k_B T \Delta
\nu_{12}}{2 \eta (g \mu_B)^2} \mu_0 \sigma_{12}^2
I_{ratio} \nu_{12}+ \dfrac{2 \tau_1}{3 h \nu_{12} V_{eff}} \dfrac{2 \Delta N_{23}+5\Delta
N_{13}}{\Delta N_{13}(\Delta N_{23}-\Delta N_{12})}
\dfrac{Power}{\xi} \right]
\end{equation}
\end{widetext}
The concentrations $N_{maser}$ and $N_{int}$ are the same only if the ESR is saturated. $N_{maser}$ is calculated 
from the rate equations and the Boltzmann law at saturation, and $N_{int}$ is calculated from WG mode behaviour. 
The ion inversion ratio $I_{ratio}$ is defined as follows:
\begin{equation}
I_{ratio}=\dfrac{(\Delta N_{23}-\Delta N_{12})(\Delta N_{13}+\Delta N_{23})}{\Delta N_{12}(5\Delta
N_{13}+2\Delta N_{23})}
\end{equation}
where the different $\Delta N_{ij}$ are the normalized population difference between the energy levels $i$
and $j$ and are calculated as shown by Benmessai et al. \cite{Karim2010}. The relaxation time
$\tau_1\approx 10$ ms at 8 K is the spin-lattice relaxation time. This time is considered as constant 
around $T_{inv}$. In reality it undergoes a direct relaxation process and follows $T^{-1}$ when $T\le 15$ K \cite{VanVleck1940, Orbach1962, Thorp1976}.
Here, $\beta_{out}$ is the coupling of the mode at 12.04 GHz, $Q_0$ 
is its unloaded $Q$-factor, $V_{eff}$ is the effective mode volume, $\eta$ is the filling factor. 
The mode parameters are constant over the temperature range in which we characterize the signal, 
and are summarised in Table \ref{tab:table3}.
\begin{table*}[h!!!]
\caption{\label{tab:table3} Measured Parameters for the WGH$_{17,0,0}$ mode at 8 K. The maser signal was operating on the lower doublet of the mode.}
\begin{ruledtabular}
\begin{tabular}{lccccccr}
Mode &  $T_{inv}$ (K) & $\nu_{WG} $ (GHz) & $\beta_{out}$ & $Q_0\times 10^6$ & $V_{eff}$ ($m^{3}$) & $\eta$\\ \hline
\multirow{2}{*}{WGH$_{17,0,0}$} & \multirow{2}{*}{7.8}  &  12.038136    & 0.200  &  600  & \multirow{2}{*}{$10^{-5}$}  & \multirow{2}{*}{0.99268}\\
					 &        &  12.038137    & 0.009  & -  &   & \\
\end{tabular}
\end{ruledtabular}
\end{table*}
\\

By definition, $\xi$ is the ratio\begin{scriptsize}
\begin{footnotesize}
•
\end{footnotesize}
\end{scriptsize} between the power in the unsaturated regime and saturation ($0 <
\xi \le 1$). Note that for the pump at 31.337 GHz where the maser is at saturation, $\xi$ should be equal to unity. 
From the results shown in Fig.~\ref{fig:6}, where $T_{inv}$ is different from one pump to another, it is
possible to calculate the corresponding concentration for each $T_{inv}$ from Eq. (\ref{eqn:2}). It is also possible to calculate the same 
concentration from Eq. (\ref{eqn:4}). Knowing the maser output power, the result are shown in Table \ref{tab:table1}.

\begin{table*}[h!!!]
\caption{\label{tab:table1} Summary of the calculated concentrations}
\begin{ruledtabular}
\begin{tabular}{lcccccr}
$\nu_{pump}$ (GHz) & $T_{inv}$ (K) & Power (dBm)\footnote{Measured at the output of resonator} & $N_{int}$ (ppb)\footnote{Calculated from Eq. \ref{eqn:3}} & $\zeta$ & $N_{maser}$ (ppb)\footnote{Calculated from Eq. \ref{eqn:4}}  & $\xi$ \\
\hline
31.320 &  9.834 & -62.59 &  13.46 & 0.0135 & 7.74 & 0.6180 \\
31.337 &  8.656 & -60.50 &  7.11   & 0.0071 & 11.09 & 1.0000 \\
31.339 &  8.169 & -65.50 &  5.32   & 0.0053 & 3.57   &  0.3162\\
31.347 &	-       &  -69.00\footnote{at 9 K}        &  -      & -        &  1.90&  0.1412          \\
\end{tabular}
\end{ruledtabular}
\end{table*}

Eq. (\ref{eqn:4}) is evaluated when the maser signal is maximum (i.e. at saturation). 
As the maser signal is oscillating at one mode (WGH$_{17,0,0}$) for all the pumps, it is clear that 
for each pump the amount of active ions is different, leading to a calculation of 
different concentrations. Each pump has a different configuration of coupling and field distribution 
in the crystal. So, each pump will interact with a different quantity of ions, and whether saturation is 
reached is determined by the coupling.


\subsection{Behavior of C2 with high concentration of Fe$^{3+}$}

It has been shown previously\cite{Creedon2009IEEE} that the concentration of Fe$^{3+}$ ions in sapphire can be increased significantly by annealing in air, causing conversion of Fe$^{2+}$ impurities to Fe$^{3+}$. Originally the two crystals C1 and C2 exhibited similar properties of very low concentration of Fe$^{3+}$ impurities. However, C2 went through a series of additional annealing and was thus transformed from the low concentration regime (as reported here for C1) to the high concentration regime. The improvement in ion concentration is permanent, and once in the high concentration regime the resonator will exhibit the effects reported in this paper. Thus, annealing a resonator is beneficial in two respects; due to the increased active ion population it will exhibit a higher maser output power, leading to a reduction in the Schawlow-Townes thermal noise limit\cite{Benmessai2008PRL}, as well as leading to a potential improvement in the curvature of the frequency-temperature dependence as is reported in this section.

For the crystal C2, the maser operation is very different as the concentration of the active 
ions is higher, resulting in a higher maser output power of $-40$ dBm, compared to $-56$ dBm 
for C1. In addition, maser signals are observed not only at the WGH$_{17,0,0}$ 
frequency, but also at different modes whose frequencies are within the ESR bandwidth 
\cite{Creedon2009IEEE}. Like C1, all the pump modes for which a maser signal can 
be excited in C2 have a temperature turnover point independent of the applied power.
Characterizing the turnover point for this crystal reveals more complex effects than 
C1. 
\begin{figure}[h!]
\centering
\includegraphics[width=0.45\textwidth]{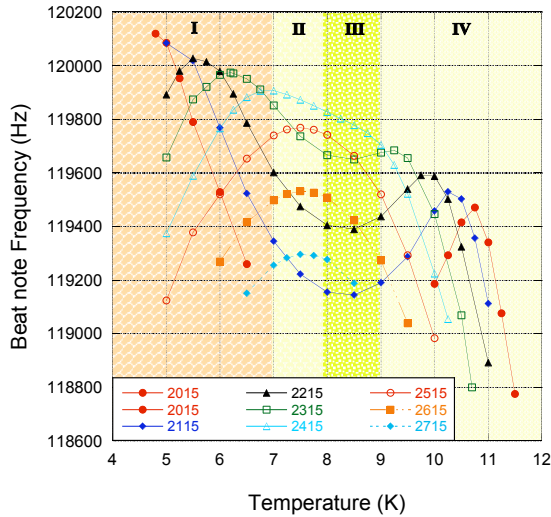}
\caption{\label{fig:7} \it{\small{Evolution of the maser signal frequency at 12.037 GHz with temperature for
different value of the same pump at offsets (in Hz) from 31.349,780,000 GHz.}}}
\end{figure}

Figure \ref{fig:7} shows the signal at 12.037 GHz oscillating for a pump applied 
at 31.349 GHz. Each curve represents the maser frequency dependence as a function of temperature for different 
pump frequencies. Here, the maser frequency shows a strong dependence on the pump tuning. 
We note that when the frequency $\nu_{pump}\leq$ 31.349,782,315 GHz, there are three
$T_{inv}$ -- one near 6 K, the second around 8 K, and the third at 10.5 K. When $\nu_{pump}\geq$
31.349,782,415 GHz, there is only one $T_{inv}$ around 8 K and the 
system behaves in the same classical fashion as the first crystal.

\begin{figure}[h!]
\centering
\includegraphics[width=0.45\textwidth]{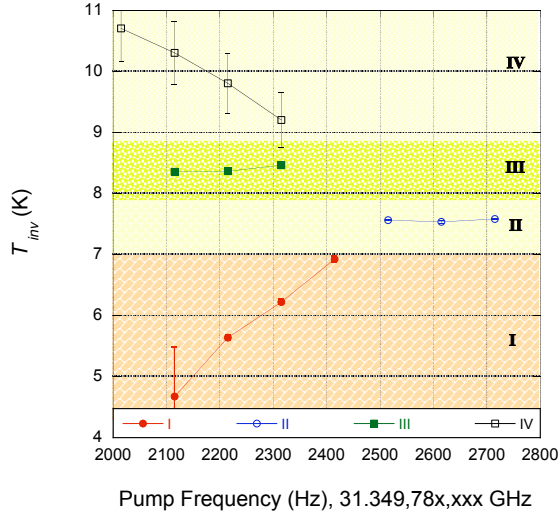}
\caption {\label{fig:8} \it{\small{Evolution of the turnover temperature at 12.037 GHz with 
different values of the same pump at offsets (in Hz) from 31.349,780,000 GHz.}}}
\end{figure}

We can also divide the figure into four zones, with each zone defining a turnover temperature.
From the results shown in Fig.~\ref{fig:7}, we can draw the evolution of the different 
turning point with the pump frequency (cf. Fig.~\ref{fig:8}).

The annulment exists for multiple temperatures only for the crystal C2 with high Fe$^{3+}$ concentration. 
This effect appears to be nonlinear, showing that a variation in the active ion population is induced as the temperature is varied, which inturn changes the turning point temperature as the temperature is varied. 

The annulment temperature depends on the distribution of the ions and the population 
difference between their levels. When the annulment temperature for the WGH$_{17,0,0}$ 
mode is measured using a network analyzer, no change is seen due to the small population difference at 
this frequency (5\% of the total active population). However, when the maser is operating, i.e. 
being pumped, the configuration of the ions is completely different: population inversion exists 
between the $\left|1/2\right\rangle$ and $\left|3/2\right\rangle$ levels, and saturation is seen at 31 GHz where the population 
of the states $\left|1/2\right\rangle$ and $\left|5/2\right\rangle$ are equal. The population inversion rate is determined by 
the choice of pump mode near 31 GHz. For each pump mode, the strength of the maser signal 
depends on its coupling and spatial field distribution in the crystal. Strong coupling allows an 
optimum exchange of energy with the ions, and field distributions occupying more volume in 
the crystal increase the interactions with different classes of ions. Considering the inhomogeneous 
broadening of the line at 31 GHz in addition to these factors, we can understand why $T_{inv}$ 
is different from one pump to another. The Fe$^{3+}$ spins act as individual packets selected 
by the pump WG mode corresponding. Thus, the pump signal selects a 
different quantity of ions among the broadened ESR frequency.

For C2, the case is more complex because $T_{inv}$ is not only different from one pump 
to another for the same maser signal, but is different for the same pump mode when the 
frequency is tuned. The maser frequency shows more than one temperature turnover point 
for some pump frequencies, and no annulment for others. In the case where there is more than one, the curvature of annulment is significantly reduced, which has the potential to improve the stability and reduce the requirements of temperature control of the maser.


\subsubsection*{Optimum Operation}
\begin{figure}[h!]
\centering
\includegraphics[width=0.45\textwidth]{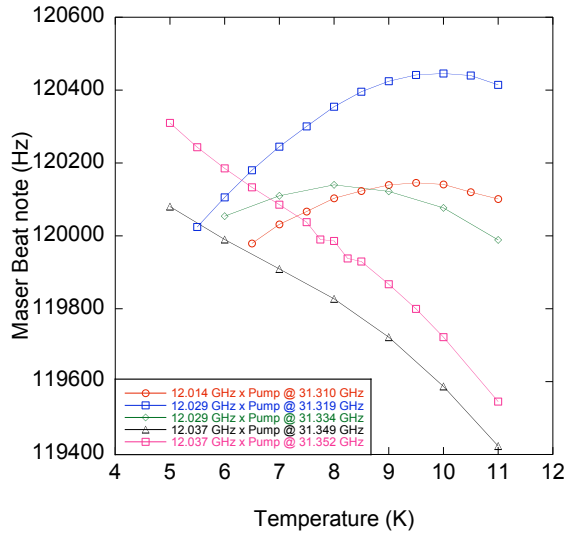}
\caption{\label{fig:9} \it{\small{Evolution of the maser signal frequency with temperature for different pumps.}}}
\end{figure}

The characterization of the maser frequency with temperature was performed at a fixed pump 
frequency. However, the pump WG mode is temperature dependent as well as the WG modes 
near 12.04 GHz. To optimize the maser operation, it is necessary to tune the pump frequency 
for each operating temperature. This optimum frequency corresponds to maxima or minima 
of the maser frequency depending on the $T_{inv}$ being measured. Fig.~\ref{fig:9} shows how 
the maser frequency varies with temperature for different pump modes.
The signal at 12.014 GHz has a turnover point at 10 K, the one at 12.029 GHz is at 10.3 K, and the one 
at 12.037 GHz shows no turnover temperature at all. The measurements were performed at different 
power levels, which were found to have no effect on the turnover temperature.


\section{Maser Temperature Operation Limit}
In order to characterize the upper temperature limit $T_{operation}$ for maser operation, we adjust the pump 
frequency for each measurement as per the last section. The results are summarised 
in Table \ref{tab:table2}. The 12.037 GHz maser signal oscillates until 28 K when pumped at 
31.349 GHz, and 30 K when pumped at 31.352 GHz. The power of the maser signal is observed to 
decrease as the temperature is raised. This is due to the fact that the populations of the Fe$^{3+}$ energy levels follow a Boltzmann distribution at thermal equilibrium. At higher temperatures, the difference in population between levels is reduced, resulting in less 
efficient masing.
At high temperature (20-30 K) the number of ions pumped at 31.3 GHz is lower than at 4.2 K, 
leading to a smaller population inversion ratio. This directly affects the maser power and 
causes it to drop as temperature rises. In addition, the absorption at 31.3 GHz is a result of 
a mixing between the $\left|1/2\right\rangle$ and $\left|5/2\right\rangle$ spin states of the Fe$^{3+}$ ion, which occurs with very small probability 
\cite{Symmons1962, Kornienko1961, Bogle1959}. The population difference at 31.3 GHz is high 
enough to excite the maser signal with this probability at low temperatures ($<$30 K), but not at high
temperatures ($>$30 K).
\begin{table}[h!]
\caption{\label{tab:table2} Different annulment and operation limit temperatures for optimal operation of the maser}
\begin{ruledtabular}
\begin{tabular}{lccr}
$\nu_{pump}$ (GHz) & $\nu_{maser} $ (GHz) & $T_{inv}$ (K) & $T_{operation}$ (K) \\
\hline
31.310 & 12.014 & 9.407 &15 \\
31.319 & 12.029 & 9.910 &26 \\
31.334 & 12.029 & 8.140 &23 \\
31.349 & 12.037 & - & 28 \\
31.352 & 12.037 & - & 30 \\
\end{tabular}
\end{ruledtabular}
\end{table}
At higher temperatures still, characterization of the WG modes around 12.04 GHz showed 
absorption effects up to 140 K. The strength of the absorption decreases when the temperature is increased, however no maser operation has been observed at these temperatures.


\section*{Conclusion}
We have demonstrated how Fe$^{3+}$ ions in a sapphire maser influence the frequency-temperature
turnover point of the maser signal. We
compared the behaviour of two crystals, one with a low active ion concentration and the other with a high
concentration. The first crystal showed
classical behaviour where $T_{inv}$ is independent of the pump power and frequency, but was different
between pump modes due to different packets
of ions being excited for each pump. The population of ions that creates the maser causes a change in the 
intrinsic annulment temperature $T_{inv}$ for the mode. The crystal with
higher Fe$^{3+}$ concentration showed a more complex behaviour, where $T_{inv}$ was different not only from
one pump mode to another, but showed
many turnover points for some choices of pump, and in some cases none at all. Thus, the curvature of the annulment point in some cases could be reduced, which would also reduce the requirements for temperature control when generating a stable frequency.


\begin{acknowledgments}
The authors wish to thank the Australian Research
Council for supporting this work under grant numbers
FL0992016, CE11E0082 and DP1092690, and
ISL grant number FR100013.
\end{acknowledgments}


\nocite{*}
\bibliography{Ref}

\providecommand{\noopsort}[1]{}\providecommand{\singleletter}[1]{#1}%
\begin{thebibliography}{32}%
\makeatletter
\providecommand \@ifxundefined [1]{%
 \@ifx{#1\undefined}
}%
\providecommand \@ifnum [1]{%
 \ifnum #1\expandafter \@firstoftwo
 \else \expandafter \@secondoftwo
 \fi
}%
\providecommand \@ifx [1]{%
 \ifx #1\expandafter \@firstoftwo
 \else \expandafter \@secondoftwo
 \fi
}%
\providecommand \natexlab [1]{#1}%
\providecommand \enquote  [1]{``#1''}%
\providecommand \bibnamefont  [1]{#1}%
\providecommand \bibfnamefont [1]{#1}%
\providecommand \citenamefont [1]{#1}%
\providecommand \href@noop [0]{\@secondoftwo}%
\providecommand \href [0]{\begingroup \@sanitize@url \@href}%
\providecommand \@href[1]{\@@startlink{#1}\@@href}%
\providecommand \@@href[1]{\endgroup#1\@@endlink}%
\providecommand \@sanitize@url [0]{\catcode `\\12\catcode `\$12\catcode
  `\&12\catcode `\#12\catcode `\^12\catcode `\_12\catcode `\%12\relax}%
\providecommand \@@startlink[1]{}%
\providecommand \@@endlink[0]{}%
\providecommand \url  [0]{\begingroup\@sanitize@url \@url }%
\providecommand \@url [1]{\endgroup\@href {#1}{\urlprefix }}%
\providecommand \urlprefix  [0]{URL }%
\providecommand \Eprint [0]{\href }%
\providecommand \doibase [0]{http://dx.doi.org/}%
\providecommand \selectlanguage [0]{\@gobble}%
\providecommand \bibinfo  [0]{\@secondoftwo}%
\providecommand \bibfield  [0]{\@secondoftwo}%
\providecommand \translation [1]{[#1]}%
\providecommand \BibitemOpen [0]{}%
\providecommand \bibitemStop [0]{}%
\providecommand \bibitemNoStop [0]{.\EOS\space}%
\providecommand \EOS [0]{\spacefactor3000\relax}%
\providecommand \BibitemShut  [1]{\csname bibitem#1\endcsname}%
\let\auto@bib@innerbib\@empty
\bibitem [{\citenamefont {Bourgeois}\ \emph {et~al.}(2005)\citenamefont
  {Bourgeois}, \citenamefont {Bazin}, \citenamefont {Kersal\'e}, \citenamefont
  {Giordano}, \citenamefont {Tobar},\ and\ \citenamefont
  {Oxborrow}}]{Pyb2005APL}%
  \BibitemOpen
  \bibfield  {author} {\bibinfo {author} {\bibfnamefont {P.-Y.}\ \bibnamefont
  {Bourgeois}}, \bibinfo {author} {\bibfnamefont {N.}~\bibnamefont {Bazin}},
  \bibinfo {author} {\bibfnamefont {Y.}~\bibnamefont {Kersal\'e}}, \bibinfo
  {author} {\bibfnamefont {V.}~\bibnamefont {Giordano}}, \bibinfo {author}
  {\bibfnamefont {M.~E.}\ \bibnamefont {Tobar}}, \ and\ \bibinfo {author}
  {\bibfnamefont {M.}~\bibnamefont {Oxborrow}},\ }\bibfield  {title} {\enquote
  {\bibinfo {title} {Maser oscillation in a whispering gallery mode microwave
  resonator},}\ }\href@noop {} {\bibfield  {journal} {\bibinfo  {journal}
  {Applied Physics Letters}\ }\textbf {\bibinfo {volume} {87}},\ \bibinfo
  {pages} {224104} (\bibinfo {year} {2005})}\BibitemShut {NoStop}%
\bibitem [{\citenamefont {Bourgeois}\ \emph {et~al.}(2006)\citenamefont
  {Bourgeois}, \citenamefont {Oxborrow}, \citenamefont {Tobar}, \citenamefont
  {Bazin}, \citenamefont {Kersal\'e},\ and\ \citenamefont
  {Giordano}}]{Pyb2006IJMP}%
  \BibitemOpen
  \bibfield  {author} {\bibinfo {author} {\bibfnamefont {P.-Y.}\ \bibnamefont
  {Bourgeois}}, \bibinfo {author} {\bibfnamefont {M.}~\bibnamefont {Oxborrow}},
  \bibinfo {author} {\bibfnamefont {M.~E.}\ \bibnamefont {Tobar}}, \bibinfo
  {author} {\bibfnamefont {N.}~\bibnamefont {Bazin}}, \bibinfo {author}
  {\bibfnamefont {Y.}~\bibnamefont {Kersal\'e}}, \ and\ \bibinfo {author}
  {\bibfnamefont {V.}~\bibnamefont {Giordano}},\ }\bibfield  {title} {\enquote
  {\bibinfo {title} {Maser oscillation from electronic spin resonance in a
  cryogenic sapphire frequency standard},}\ }\href@noop {} {\bibfield
  {journal} {\bibinfo  {journal} {International Journal of Modern Physics}\
  }\textbf {\bibinfo {volume} {B 20}},\ \bibinfo {pages} {1606} (\bibinfo
  {year} {2006})}\BibitemShut {NoStop}%
\bibitem [{\citenamefont {Benmessai}\ \emph {et~al.}(2005)\citenamefont
  {Benmessai}, \citenamefont {Bourgeois}, \citenamefont {Kersal\'e},
  \citenamefont {Bazin}, \citenamefont {Tobar}, \citenamefont {Hartnett},
  \citenamefont {Oxborrow},\ and\ \citenamefont {Giordano}}]{Benmessai2005EL}%
  \BibitemOpen
  \bibfield  {author} {\bibinfo {author} {\bibfnamefont {K.}~\bibnamefont
  {Benmessai}}, \bibinfo {author} {\bibfnamefont {P.-Y.}\ \bibnamefont
  {Bourgeois}}, \bibinfo {author} {\bibfnamefont {Y.}~\bibnamefont
  {Kersal\'e}}, \bibinfo {author} {\bibfnamefont {N.}~\bibnamefont {Bazin}},
  \bibinfo {author} {\bibfnamefont {M.~E.}\ \bibnamefont {Tobar}}, \bibinfo
  {author} {\bibfnamefont {J.~G.}\ \bibnamefont {Hartnett}}, \bibinfo {author}
  {\bibfnamefont {M.}~\bibnamefont {Oxborrow}}, \ and\ \bibinfo {author}
  {\bibfnamefont {V.}~\bibnamefont {Giordano}},\ }\bibfield  {title} {\enquote
  {\bibinfo {title} {Frequency instability measurement system of cryogenic
  maser oscillator},}\ }\href@noop {} {\bibfield  {journal} {\bibinfo
  {journal} {Electronics Letters}\ }\textbf {\bibinfo {volume} {43}},\ \bibinfo
  {pages} {1436} (\bibinfo {year} {2005})}\BibitemShut {NoStop}%
\bibitem [{\citenamefont {Benmessai}\ \emph {et~al.}(2008)\citenamefont
  {Benmessai}, \citenamefont {Creedon}, \citenamefont {Tobar}, \citenamefont
  {Bourgeois}, \citenamefont {Kersal\'e},\ and\ \citenamefont
  {Giordano}}]{Benmessai2008PRL}%
  \BibitemOpen
  \bibfield  {author} {\bibinfo {author} {\bibfnamefont {K.}~\bibnamefont
  {Benmessai}}, \bibinfo {author} {\bibfnamefont {D.~L.}\ \bibnamefont
  {Creedon}}, \bibinfo {author} {\bibfnamefont {M.~E.}\ \bibnamefont {Tobar}},
  \bibinfo {author} {\bibfnamefont {P.-Y.}\ \bibnamefont {Bourgeois}}, \bibinfo
  {author} {\bibfnamefont {Y.}~\bibnamefont {Kersal\'e}}, \ and\ \bibinfo
  {author} {\bibfnamefont {V.}~\bibnamefont {Giordano}},\ }\bibfield  {title}
  {\enquote {\bibinfo {title} {Measurement of the fundamental thermal noise
  limit in a cryogenic sapphire frequency standard using bimodal maser
  oscillations},}\ }\href@noop {} {\bibfield  {journal} {\bibinfo  {journal}
  {Physical Review Letters}\ }\textbf {\bibinfo {volume} {100}},\ \bibinfo
  {pages} {233901} (\bibinfo {year} {2008})}\BibitemShut {NoStop}%
\bibitem [{\citenamefont {Creedon}\ \emph {et~al.}(2009)\citenamefont
  {Creedon}, \citenamefont {Benmessai}, \citenamefont {Tobar}, \citenamefont
  {Hartnett}, \citenamefont {Bourgeois}, \citenamefont {Kersal\'e},
  \citenamefont {LeFloch},\ and\ \citenamefont {Giordano}}]{Creedon2009IEEE}%
  \BibitemOpen
  \bibfield  {author} {\bibinfo {author} {\bibfnamefont {D.~L.}\ \bibnamefont
  {Creedon}}, \bibinfo {author} {\bibfnamefont {K.}~\bibnamefont {Benmessai}},
  \bibinfo {author} {\bibfnamefont {M.~E.}\ \bibnamefont {Tobar}}, \bibinfo
  {author} {\bibfnamefont {J.}~\bibnamefont {Hartnett}}, \bibinfo {author}
  {\bibfnamefont {P.-Y.}\ \bibnamefont {Bourgeois}}, \bibinfo {author}
  {\bibfnamefont {Y.}~\bibnamefont {Kersal\'e}}, \bibinfo {author}
  {\bibfnamefont {J.-M.}\ \bibnamefont {LeFloch}}, \ and\ \bibinfo {author}
  {\bibfnamefont {V.}~\bibnamefont {Giordano}},\ }\bibfield  {title} {\enquote
  {\bibinfo {title} {High power solid-state sapphire whispering gallery mode
  maser},}\ }\href@noop {} {\bibfield  {journal} {\bibinfo  {journal} {IEEE
  Transactions on Ultrasonics, Ferroelectrics and Frequency Control}\ }\textbf
  {\bibinfo {volume} {57}},\ \bibinfo {pages} {282--285} (\bibinfo {year}
  {2009})}\BibitemShut {NoStop}%
\bibitem [{\citenamefont {Kovacich}, \citenamefont {Mann},\ and\ \citenamefont
  {Blair}(1997)}]{Kovacich1997JPDAP}%
  \BibitemOpen
  \bibfield  {author} {\bibinfo {author} {\bibfnamefont {R.~P.}\ \bibnamefont
  {Kovacich}}, \bibinfo {author} {\bibfnamefont {A.~G.}\ \bibnamefont {Mann}},
  \ and\ \bibinfo {author} {\bibfnamefont {D.~G.}\ \bibnamefont {Blair}},\
  }\bibfield  {title} {\enquote {\bibinfo {title} {Magnetic field tuning of
  paramagnetic frequency - temperature compensation in cryogenic sapphire
  dielectric microwave resonators},}\ }\href@noop {} {\bibfield  {journal}
  {\bibinfo  {journal} {Journal of Physics D: Applied Physics}\ }\textbf
  {\bibinfo {volume} {30}},\ \bibinfo {pages} {3146--3152} (\bibinfo {year}
  {1997})}\BibitemShut {NoStop}%
\bibitem [{\citenamefont {Mann}\ \emph {et~al.}(1991)\citenamefont {Mann},
  \citenamefont {Giles}, \citenamefont {Blair},\ and\ \citenamefont
  {Buckingham}}]{Giles1991JPDAP}%
  \BibitemOpen
  \bibfield  {author} {\bibinfo {author} {\bibfnamefont {A.~G.}\ \bibnamefont
  {Mann}}, \bibinfo {author} {\bibfnamefont {A.~J.}\ \bibnamefont {Giles}},
  \bibinfo {author} {\bibfnamefont {D.~G.}\ \bibnamefont {Blair}}, \ and\
  \bibinfo {author} {\bibfnamefont {M.~J.}\ \bibnamefont {Buckingham}},\
  }\bibfield  {title} {\enquote {\bibinfo {title} {Ultra-stable cryogenic
  sapphire dielectric resonators: mode frequency-temperature compensation by
  residual paramagnetic impurities},}\ }\href@noop {} {\bibfield  {journal}
  {\bibinfo  {journal} {Journal of Physics D: Applied Physics}\ }\textbf
  {\bibinfo {volume} {25}},\ \bibinfo {pages} {1105--1109} (\bibinfo {year}
  {1991})}\BibitemShut {NoStop}%
\bibitem [{\citenamefont {Hartnett}\ \emph {et~al.}(1991)\citenamefont
  {Hartnett}, \citenamefont {Tobar}, \citenamefont {LeFloch}, \citenamefont
  {Krupka},\ and\ \citenamefont {Bourgeois}}]{Hartenett2007PRB}%
  \BibitemOpen
  \bibfield  {author} {\bibinfo {author} {\bibfnamefont {J.~G.}\ \bibnamefont
  {Hartnett}}, \bibinfo {author} {\bibfnamefont {M.~E.}\ \bibnamefont {Tobar}},
  \bibinfo {author} {\bibfnamefont {J.-M.}\ \bibnamefont {LeFloch}}, \bibinfo
  {author} {\bibfnamefont {J.}~\bibnamefont {Krupka}}, \ and\ \bibinfo {author}
  {\bibfnamefont {P.-Y.}\ \bibnamefont {Bourgeois}},\ }\bibfield  {title}
  {\enquote {\bibinfo {title} {Anisotropic paramagnetic susceptibility of
  crystalline ruby at cryogenic temperatures},}\ }\href@noop {} {\bibfield
  {journal} {\bibinfo  {journal} {Physical Review B}\ }\textbf {\bibinfo
  {volume} {75}},\ \bibinfo {pages} {024415} (\bibinfo {year}
  {1991})}\BibitemShut {NoStop}%
\bibitem [{\citenamefont {Dick}, \citenamefont {Santiago},\ and\ \citenamefont
  {Wang}(1995)}]{Dick1995IEEE}%
  \BibitemOpen
  \bibfield  {author} {\bibinfo {author} {\bibfnamefont {G.~J.}\ \bibnamefont
  {Dick}}, \bibinfo {author} {\bibfnamefont {D.~G.}\ \bibnamefont {Santiago}},
  \ and\ \bibinfo {author} {\bibfnamefont {R.~T.}\ \bibnamefont {Wang}},\
  }\bibfield  {title} {\enquote {\bibinfo {title} {Temperature-compensated
  sapphire resonator for ultra-stable oscillator capability at temperatures
  above 77{K}},}\ }\href@noop {} {\bibfield  {journal} {\bibinfo  {journal}
  {IEEE Transactions on Ultrasonics, Ferroelectrics, and Frequency Control}\
  }\textbf {\bibinfo {volume} {5}},\ \bibinfo {pages} {812--819} (\bibinfo
  {year} {1995})}\BibitemShut {NoStop}%
\bibitem [{\citenamefont {Dick}, \citenamefont {Wang},\ and\ \citenamefont
  {Tjoelker}(1998)}]{Dick1998IFCS}%
  \BibitemOpen
  \bibfield  {author} {\bibinfo {author} {\bibfnamefont {G.~J.}\ \bibnamefont
  {Dick}}, \bibinfo {author} {\bibfnamefont {R.~T.}\ \bibnamefont {Wang}}, \
  and\ \bibinfo {author} {\bibfnamefont {R.~L.}\ \bibnamefont {Tjoelker}},\
  }\bibfield  {title} {\enquote {\bibinfo {title} {Cryo-cooled sapphire
  oscillator with ultra-high stability},}\ }in\ \href@noop {} {\emph {\bibinfo
  {booktitle} {Proceedings of the IEEE International Frequency Control
  Symposium and Exposition}}}\ (\bibinfo {year} {1998})\ pp.\ \bibinfo {pages}
  {528--533}\BibitemShut {NoStop}%
\bibitem [{\citenamefont {Wang}\ and\ \citenamefont
  {Dick}(2002)}]{Wang2002IEEE}%
  \BibitemOpen
  \bibfield  {author} {\bibinfo {author} {\bibfnamefont {R.~T.}\ \bibnamefont
  {Wang}}\ and\ \bibinfo {author} {\bibfnamefont {G.~J.}\ \bibnamefont
  {Dick}},\ }\bibfield  {title} {\enquote {\bibinfo {title} {Cryo-cooled
  sapphire oscillator with mechanical compensation},}\ }\href@noop {}
  {\bibfield  {journal} {\bibinfo  {journal} {IEEE Transactions on Ultrasonics,
  Ferroelectrics, and Frequency Control}\ ,\ \bibinfo {pages} {543--547}}
  (\bibinfo {year} {2002})}\BibitemShut {NoStop}%
\bibitem [{\citenamefont {Boubekeur}\ \emph {et~al.}(2005)\citenamefont
  {Boubekeur}, \citenamefont {Hartnett}, \citenamefont {Tobar}, \citenamefont
  {Bazin}, \citenamefont {Kersal\'{e}},\ and\ \citenamefont
  {Giordano}}]{Naimi2005}%
  \BibitemOpen
  \bibfield  {author} {\bibinfo {author} {\bibfnamefont {N.}~\bibnamefont
  {Boubekeur}}, \bibinfo {author} {\bibfnamefont {J.~G.}\ \bibnamefont
  {Hartnett}}, \bibinfo {author} {\bibfnamefont {M.~E.}\ \bibnamefont {Tobar}},
  \bibinfo {author} {\bibfnamefont {N.}~\bibnamefont {Bazin}}, \bibinfo
  {author} {\bibfnamefont {Y.}~\bibnamefont {Kersal\'{e}}}, \ and\ \bibinfo
  {author} {\bibfnamefont {V.}~\bibnamefont {Giordano}},\ }\bibfield  {title}
  {\enquote {\bibinfo {title} {Frequency stability of {Ti$^{3+}$}-doped
  whispering gallery mode sapphire resonator oscillator at 34{K}},}\
  }\href@noop {} {\bibfield  {journal} {\bibinfo  {journal} {Electronics
  Letters}\ }\textbf {\bibinfo {volume} {9}},\ \bibinfo {pages} {534--535}
  (\bibinfo {year} {2005})}\BibitemShut {NoStop}%
\bibitem [{\citenamefont {Tobar}\ \emph {et~al.}(1999)\citenamefont {Tobar},
  \citenamefont {Hartnett}, \citenamefont {Cros}, \citenamefont {Blondy},
  \citenamefont {Krupka}, \citenamefont {Ivanov},\ and\ \citenamefont
  {Guillon}}]{tobar1999}%
  \BibitemOpen
  \bibfield  {author} {\bibinfo {author} {\bibfnamefont {M.~E.}\ \bibnamefont
  {Tobar}}, \bibinfo {author} {\bibfnamefont {J.~G.}\ \bibnamefont {Hartnett}},
  \bibinfo {author} {\bibfnamefont {D.}~\bibnamefont {Cros}}, \bibinfo {author}
  {\bibfnamefont {P.}~\bibnamefont {Blondy}}, \bibinfo {author} {\bibfnamefont
  {J.}~\bibnamefont {Krupka}}, \bibinfo {author} {\bibfnamefont {E.~N.}\
  \bibnamefont {Ivanov}}, \ and\ \bibinfo {author} {\bibfnamefont
  {P.}~\bibnamefont {Guillon}},\ }\bibfield  {title} {\enquote {\bibinfo
  {title} {Design of high-{Q} frequency-temperature compensated dielectric
  resonators},}\ }\href@noop {} {\bibfield  {journal} {\bibinfo  {journal}
  {Electronics Letters.}\ }\textbf {\bibinfo {volume} {4}},\ \bibinfo {pages}
  {303--305} (\bibinfo {year} {1999})}\BibitemShut {NoStop}%
\bibitem [{\citenamefont {Creedon}\ \emph {et~al.}(2010)\citenamefont
  {Creedon}, \citenamefont {Tobar}, \citenamefont {LeFloch}, \citenamefont
  {Reshitnky},\ and\ \citenamefont {Duty}}]{Creedon2010PRB}%
  \BibitemOpen
  \bibfield  {author} {\bibinfo {author} {\bibfnamefont {D.~L.}\ \bibnamefont
  {Creedon}}, \bibinfo {author} {\bibfnamefont {M.~E.}\ \bibnamefont {Tobar}},
  \bibinfo {author} {\bibfnamefont {J.~M.}\ \bibnamefont {LeFloch}}, \bibinfo
  {author} {\bibfnamefont {Y.}~\bibnamefont {Reshitnky}}, \ and\ \bibinfo
  {author} {\bibfnamefont {T.}~\bibnamefont {Duty}},\ }\bibfield  {title}
  {\enquote {\bibinfo {title} {Single-crystal sapphire resonator at millikelvin
  temperatures: Observation of thermal bistability in high-{Q} factor
  whispering gallery modes},}\ }\href@noop {} {\bibfield  {journal} {\bibinfo
  {journal} {Physical Review B}\ }\textbf {\bibinfo {volume} {82}},\ \bibinfo
  {pages} {104305} (\bibinfo {year} {2010})}\BibitemShut {NoStop}%
\bibitem [{\citenamefont {Creedon}\ \emph {et~al.}(2011)\citenamefont
  {Creedon}, \citenamefont {Reshitnky}, \citenamefont {Farr}, \citenamefont
  {Martinis}, \citenamefont {Duty},\ and\ \citenamefont
  {Tobar}}]{Creedon2011APL}%
  \BibitemOpen
  \bibfield  {author} {\bibinfo {author} {\bibfnamefont {D.~L.}\ \bibnamefont
  {Creedon}}, \bibinfo {author} {\bibfnamefont {Y.}~\bibnamefont {Reshitnky}},
  \bibinfo {author} {\bibfnamefont {Y.}~\bibnamefont {Farr}}, \bibinfo {author}
  {\bibfnamefont {W.}~\bibnamefont {Martinis}}, \bibinfo {author}
  {\bibfnamefont {T.}~\bibnamefont {Duty}}, \ and\ \bibinfo {author}
  {\bibfnamefont {M.~E.}\ \bibnamefont {Tobar}},\ }\bibfield  {title} {\enquote
  {\bibinfo {title} {High {Q}-factor sapphire whispering gallery mode microwave
  resonator at single photon energies and milli-kelvin temperatures},}\
  }\href@noop {} {\bibfield  {journal} {\bibinfo  {journal} {Applied Physics
  letters}\ }\textbf {\bibinfo {volume} {98}},\ \bibinfo {pages} {222903}
  (\bibinfo {year} {2011})}\BibitemShut {NoStop}%
\bibitem [{\citenamefont {Duty}(2010)}]{Tim}%
  \BibitemOpen
  \bibfield  {author} {\bibinfo {author} {\bibfnamefont {T.}~\bibnamefont
  {Duty}},\ }\bibfield  {title} {\enquote {\bibinfo {title} {Towards
  superconductor-spin ensemble hybrid quantum systems},}\ }\href@noop {}
  {\bibfield  {journal} {\bibinfo  {journal} {Physics}\ }\textbf {\bibinfo
  {volume} {3}} (\bibinfo {year} {2010})}\BibitemShut {NoStop}%
\bibitem [{\citenamefont {Jelezko}\ \emph {et~al.}(2004)\citenamefont
  {Jelezko}, \citenamefont {Gaebel}, \citenamefont {Popa}, \citenamefont
  {Gruber},\ and\ \citenamefont {Wrachtrup}}]{PhysRevLett.92.076401}%
  \BibitemOpen
  \bibfield  {author} {\bibinfo {author} {\bibfnamefont {F.}~\bibnamefont
  {Jelezko}}, \bibinfo {author} {\bibfnamefont {T.}~\bibnamefont {Gaebel}},
  \bibinfo {author} {\bibfnamefont {I.}~\bibnamefont {Popa}}, \bibinfo {author}
  {\bibfnamefont {A.}~\bibnamefont {Gruber}}, \ and\ \bibinfo {author}
  {\bibfnamefont {J.}~\bibnamefont {Wrachtrup}},\ }\bibfield  {title} {\enquote
  {\bibinfo {title} {Observation of coherent oscillations in a single electron
  spin},}\ }\href@noop {} {\bibfield  {journal} {\bibinfo  {journal} {Physical
  Review Letters}\ }\textbf {\bibinfo {volume} {92}},\ \bibinfo {pages}
  {076401} (\bibinfo {year} {2004})}\BibitemShut {NoStop}%
\bibitem [{\citenamefont {Kubo}\ \emph {et~al.}(2010)\citenamefont {Kubo},
  \citenamefont {Ong}, \citenamefont {Bertet}, \citenamefont {Vion},
  \citenamefont {Jacques}, \citenamefont {Zheng}, \citenamefont {Dr\'eau},
  \citenamefont {Roch}, \citenamefont {Auffeves}, \citenamefont {Jelezko},
  \citenamefont {Wrachtrup}, \citenamefont {Barthe}, \citenamefont {Bergonzo},\
  and\ \citenamefont {Esteve}}]{PhysRevLett.105.140502}%
  \BibitemOpen
  \bibfield  {author} {\bibinfo {author} {\bibfnamefont {Y.}~\bibnamefont
  {Kubo}}, \bibinfo {author} {\bibfnamefont {F.~R.}\ \bibnamefont {Ong}},
  \bibinfo {author} {\bibfnamefont {P.}~\bibnamefont {Bertet}}, \bibinfo
  {author} {\bibfnamefont {D.}~\bibnamefont {Vion}}, \bibinfo {author}
  {\bibfnamefont {V.}~\bibnamefont {Jacques}}, \bibinfo {author} {\bibfnamefont
  {D.}~\bibnamefont {Zheng}}, \bibinfo {author} {\bibfnamefont
  {A.}~\bibnamefont {Dr\'eau}}, \bibinfo {author} {\bibfnamefont {J.-F.}\
  \bibnamefont {Roch}}, \bibinfo {author} {\bibfnamefont {A.}~\bibnamefont
  {Auffeves}}, \bibinfo {author} {\bibfnamefont {F.}~\bibnamefont {Jelezko}},
  \bibinfo {author} {\bibfnamefont {J.}~\bibnamefont {Wrachtrup}}, \bibinfo
  {author} {\bibfnamefont {M.~F.}\ \bibnamefont {Barthe}}, \bibinfo {author}
  {\bibfnamefont {P.}~\bibnamefont {Bergonzo}}, \ and\ \bibinfo {author}
  {\bibfnamefont {D.}~\bibnamefont {Esteve}},\ }\bibfield  {title} {\enquote
  {\bibinfo {title} {Strong coupling of a spin ensemble to a superconducting
  resonator},}\ }\href@noop {} {\bibfield  {journal} {\bibinfo  {journal}
  {Physical Review Letters}\ }\textbf {\bibinfo {volume} {105}},\ \bibinfo
  {pages} {140502} (\bibinfo {year} {2010})}\BibitemShut {NoStop}%
\bibitem [{\citenamefont {Schuster}\ \emph {et~al.}(2010)\citenamefont
  {Schuster}, \citenamefont {Sears}, \citenamefont {Ginossar}, \citenamefont
  {DiCarlo}, \citenamefont {Frunzio}, \citenamefont {Morton}, \citenamefont
  {Wu}, \citenamefont {Briggs}, \citenamefont {Buckley}, \citenamefont
  {Awschalom},\ and\ \citenamefont {Schoelkopf}}]{PhysRevLett.105.140501}%
  \BibitemOpen
  \bibfield  {author} {\bibinfo {author} {\bibfnamefont {D.~I.}\ \bibnamefont
  {Schuster}}, \bibinfo {author} {\bibfnamefont {A.~P.}\ \bibnamefont {Sears}},
  \bibinfo {author} {\bibfnamefont {E.}~\bibnamefont {Ginossar}}, \bibinfo
  {author} {\bibfnamefont {L.}~\bibnamefont {DiCarlo}}, \bibinfo {author}
  {\bibfnamefont {L.}~\bibnamefont {Frunzio}}, \bibinfo {author} {\bibfnamefont
  {J.~J.~L.}\ \bibnamefont {Morton}}, \bibinfo {author} {\bibfnamefont
  {H.}~\bibnamefont {Wu}}, \bibinfo {author} {\bibfnamefont {G.~A.~D.}\
  \bibnamefont {Briggs}}, \bibinfo {author} {\bibfnamefont {B.~B.}\
  \bibnamefont {Buckley}}, \bibinfo {author} {\bibfnamefont {D.~D.}\
  \bibnamefont {Awschalom}}, \ and\ \bibinfo {author} {\bibfnamefont {R.~J.}\
  \bibnamefont {Schoelkopf}},\ }\bibfield  {title} {\enquote {\bibinfo {title}
  {High-cooperativity coupling of electron-spin ensembles to superconducting
  cavities},}\ }\href@noop {} {\bibfield  {journal} {\bibinfo  {journal}
  {Physical Review Letters}\ }\textbf {\bibinfo {volume} {105}},\ \bibinfo
  {pages} {140501} (\bibinfo {year} {2010})}\BibitemShut {NoStop}%
\bibitem [{\citenamefont {Fuchs}\ \emph {et~al.}(2010)\citenamefont {Fuchs},
  \citenamefont {Dobrovitski}, \citenamefont {Toyli}, \citenamefont {Heremans},
  \citenamefont {Weis}, \citenamefont {Schenkel},\ and\ \citenamefont
  {Awschalom}}]{FuchsNatPhys}%
  \BibitemOpen
  \bibfield  {author} {\bibinfo {author} {\bibfnamefont {G.~D.}\ \bibnamefont
  {Fuchs}}, \bibinfo {author} {\bibfnamefont {V.~V.}\ \bibnamefont
  {Dobrovitski}}, \bibinfo {author} {\bibfnamefont {D.~M.}\ \bibnamefont
  {Toyli}}, \bibinfo {author} {\bibfnamefont {F.~J.}\ \bibnamefont {Heremans}},
  \bibinfo {author} {\bibfnamefont {C.~D.}\ \bibnamefont {Weis}}, \bibinfo
  {author} {\bibfnamefont {T.}~\bibnamefont {Schenkel}}, \ and\ \bibinfo
  {author} {\bibfnamefont {D.~D.}\ \bibnamefont {Awschalom}},\ }\bibfield
  {title} {\enquote {\bibinfo {title} {Excited-state spin coherence of a single
  nitrogen-vacancy centre in diamond},}\ }\href@noop {} {\bibfield  {journal}
  {\bibinfo  {journal} {Nature Physics}\ }\textbf {\bibinfo {volume} {6}},\
  \bibinfo {pages} {668--672} (\bibinfo {year} {2010})}\BibitemShut {NoStop}%
\bibitem [{\citenamefont {Wu}\ \emph {et~al.}(2010)\citenamefont {Wu},
  \citenamefont {George}, \citenamefont {Wesenberg}, \citenamefont {M\o{}lmer},
  \citenamefont {Schuster}, \citenamefont {Schoelkopf}, \citenamefont {Itoh},
  \citenamefont {Ardavan}, \citenamefont {Morton},\ and\ \citenamefont
  {Briggs}}]{PhysRevLett.105.140503}%
  \BibitemOpen
  \bibfield  {author} {\bibinfo {author} {\bibfnamefont {H.}~\bibnamefont
  {Wu}}, \bibinfo {author} {\bibfnamefont {R.~E.}\ \bibnamefont {George}},
  \bibinfo {author} {\bibfnamefont {J.~H.}\ \bibnamefont {Wesenberg}}, \bibinfo
  {author} {\bibfnamefont {K.}~\bibnamefont {M\o{}lmer}}, \bibinfo {author}
  {\bibfnamefont {D.~I.}\ \bibnamefont {Schuster}}, \bibinfo {author}
  {\bibfnamefont {R.~J.}\ \bibnamefont {Schoelkopf}}, \bibinfo {author}
  {\bibfnamefont {K.~M.}\ \bibnamefont {Itoh}}, \bibinfo {author}
  {\bibfnamefont {A.}~\bibnamefont {Ardavan}}, \bibinfo {author} {\bibfnamefont
  {J.~J.~L.}\ \bibnamefont {Morton}}, \ and\ \bibinfo {author} {\bibfnamefont
  {G.~A.~D.}\ \bibnamefont {Briggs}},\ }\bibfield  {title} {\enquote {\bibinfo
  {title} {Storage of multiple coherent microwave excitations in an electron
  spin ensemble},}\ }\href@noop {} {\bibfield  {journal} {\bibinfo  {journal}
  {Physical Review Letters}\ }\textbf {\bibinfo {volume} {105}},\ \bibinfo
  {pages} {140503} (\bibinfo {year} {2010})}\BibitemShut {NoStop}%
\bibitem [{\citenamefont {Chiorescu}\ \emph {et~al.}(2010)\citenamefont
  {Chiorescu}, \citenamefont {Groll}, \citenamefont {Bertaina}, \citenamefont
  {Mori},\ and\ \citenamefont {Miyashita}}]{PhysRevB.82.024413}%
  \BibitemOpen
  \bibfield  {author} {\bibinfo {author} {\bibfnamefont {I.}~\bibnamefont
  {Chiorescu}}, \bibinfo {author} {\bibfnamefont {N.}~\bibnamefont {Groll}},
  \bibinfo {author} {\bibfnamefont {S.}~\bibnamefont {Bertaina}}, \bibinfo
  {author} {\bibfnamefont {T.}~\bibnamefont {Mori}}, \ and\ \bibinfo {author}
  {\bibfnamefont {S.}~\bibnamefont {Miyashita}},\ }\bibfield  {title} {\enquote
  {\bibinfo {title} {Magnetic strong coupling in a spin-photon system and
  transition to classical regime},}\ }\href@noop {} {\bibfield  {journal}
  {\bibinfo  {journal} {Physical Review B}\ }\textbf {\bibinfo {volume} {82}},\
  \bibinfo {pages} {024413} (\bibinfo {year} {2010})}\BibitemShut {NoStop}%
\bibitem [{\citenamefont {Benmessai}(2008)}]{mathese}%
  \BibitemOpen
  \bibfield  {author} {\bibinfo {author} {\bibfnamefont {K.}~\bibnamefont
  {Benmessai}},\ }\emph {\bibinfo {title} {Maser Cryog\' enique \`a Modes de
  Gallerie}},\ \href@noop {} {\bibinfo {type} {{Ph.D.} thesis}},\ \bibinfo
  {school} {Universit\' e de Franche-Comt\' e} (\bibinfo {year}
  {2008})\BibitemShut {NoStop}%
\bibitem [{\citenamefont {Bogle}\ and\ \citenamefont
  {Symmons}(1958)}]{Bogle1958}%
  \BibitemOpen
  \bibfield  {author} {\bibinfo {author} {\bibfnamefont {G.~S.}\ \bibnamefont
  {Bogle}}\ and\ \bibinfo {author} {\bibfnamefont {H.~F.}\ \bibnamefont
  {Symmons}},\ }\bibfield  {title} {\enquote {\bibinfo {title} {Zero-fieald
  masers},}\ }\href@noop {} {\bibfield  {journal} {\bibinfo  {journal}
  {Australian Journal of Physics}\ }\textbf {\bibinfo {volume} {12}},\ \bibinfo
  {pages} {1} (\bibinfo {year} {1958})}\BibitemShut {NoStop}%
\bibitem [{\citenamefont {Siegman}(1964)}]{Siegman}%
  \BibitemOpen
  \bibfield  {author} {\bibinfo {author} {\bibfnamefont {A.~E.}\ \bibnamefont
  {Siegman}},\ }\href@noop {} {\emph {\bibinfo {title} {Microwave Solid State
  Masers}}}\ (\bibinfo  {publisher} {Mc. Graw-Hill},\ \bibinfo {year}
  {1964})\BibitemShut {NoStop}%
\bibitem [{\citenamefont {Symmons}\ and\ \citenamefont
  {Bogle}(1962)}]{Symmons1962}%
  \BibitemOpen
  \bibfield  {author} {\bibinfo {author} {\bibfnamefont {H.~F.}\ \bibnamefont
  {Symmons}}\ and\ \bibinfo {author} {\bibfnamefont {G.~S.}\ \bibnamefont
  {Bogle}},\ }\bibfield  {title} {\enquote {\bibinfo {title} {On the exactness
  of the spin-{H}amiltonian description of {Fe$^{3+}$} in sapphire},}\
  }\href@noop {} {\bibfield  {journal} {\bibinfo  {journal} {Proceedings of the
  Physical Society}\ }\textbf {\bibinfo {volume} {79}},\ \bibinfo {pages}
  {468--472} (\bibinfo {year} {1962})}\BibitemShut {NoStop}%
\bibitem [{\citenamefont {Benmessai}\ \emph {et~al.}(2010)\citenamefont
  {Benmessai}, \citenamefont {Bourgeois}, \citenamefont {Tobar}, \citenamefont
  {Bazin}, \citenamefont {Kersal\'{e}},\ and\ \citenamefont
  {Giordano}}]{Karim2010}%
  \BibitemOpen
  \bibfield  {author} {\bibinfo {author} {\bibfnamefont {K.}~\bibnamefont
  {Benmessai}}, \bibinfo {author} {\bibfnamefont {P.-Y.}\ \bibnamefont
  {Bourgeois}}, \bibinfo {author} {\bibfnamefont {M.~E.}\ \bibnamefont
  {Tobar}}, \bibinfo {author} {\bibfnamefont {N.}~\bibnamefont {Bazin}},
  \bibinfo {author} {\bibfnamefont {Y.}~\bibnamefont {Kersal\'{e}}}, \ and\
  \bibinfo {author} {\bibfnamefont {V.}~\bibnamefont {Giordano}},\ }\bibfield
  {title} {\enquote {\bibinfo {title} {Amplification process in a high-{Q}
  cryogenic whispering gallery mode sapphire {F}e$^{3 +}$ maser},}\ }\href@noop
  {} {\bibfield  {journal} {\bibinfo  {journal} {Measurement Science and
  Technology}\ }\textbf {\bibinfo {volume} {21}} (\bibinfo {year}
  {2010})}\BibitemShut {NoStop}%
\bibitem [{\citenamefont {Vleck}(1940)}]{VanVleck1940}%
  \BibitemOpen
  \bibfield  {author} {\bibinfo {author} {\bibfnamefont {J.~V.}\ \bibnamefont
  {Vleck}},\ }\bibfield  {title} {\enquote {\bibinfo {title} {Paramagnetic
  relaxation times for titanium and chrome alum},}\ }\href@noop {} {\bibfield
  {journal} {\bibinfo  {journal} {Physical Review}\ }\textbf {\bibinfo {volume}
  {57}},\ \bibinfo {pages} {426--447} (\bibinfo {year} {1940})}\BibitemShut
  {NoStop}%
\bibitem [{\citenamefont {Orbach}\ and\ \citenamefont
  {Blume}(1962)}]{Orbach1962}%
  \BibitemOpen
  \bibfield  {author} {\bibinfo {author} {\bibfnamefont {R.}~\bibnamefont
  {Orbach}}\ and\ \bibinfo {author} {\bibfnamefont {M.}~\bibnamefont {Blume}},\
  }\bibfield  {title} {\enquote {\bibinfo {title} {Spin-lattice relaxation in
  multilevel spin systems},}\ }\href@noop {} {\bibfield  {journal} {\bibinfo
  {journal} {Physical Review Letters}\ }\textbf {\bibinfo {volume} {8}},\
  \bibinfo {pages} {478--480} (\bibinfo {year} {1962})}\BibitemShut {NoStop}%
\bibitem [{\citenamefont {Thorp}\ and\ \citenamefont
  {Ammar}(1976)}]{Thorp1976}%
  \BibitemOpen
  \bibfield  {author} {\bibinfo {author} {\bibfnamefont {J.}~\bibnamefont
  {Thorp}}\ and\ \bibinfo {author} {\bibfnamefont {E.~A.~E.}\ \bibnamefont
  {Ammar}},\ }\bibfield  {title} {\enquote {\bibinfo {title} {Spin-lattice
  relaxation in gadolinium-doped calcium tungstate},}\ }\href@noop {}
  {\bibfield  {journal} {\bibinfo  {journal} {Journal of Materials Science}\
  }\textbf {\bibinfo {volume} {11}},\ \bibinfo {pages} {1215--1219} (\bibinfo
  {year} {1976})}\BibitemShut {NoStop}%
\bibitem [{\citenamefont {Kornienko}\ and\ \citenamefont
  {Prokhorov}(1961)}]{Kornienko1961}%
  \BibitemOpen
  \bibfield  {author} {\bibinfo {author} {\bibfnamefont {L.~S.}\ \bibnamefont
  {Kornienko}}\ and\ \bibinfo {author} {\bibfnamefont {A.~M.}\ \bibnamefont
  {Prokhorov}},\ }\bibfield  {title} {\enquote {\bibinfo {title} {Electronic
  paramagnetic resonance of the {Fe$^{3+}$} ion in corundum},}\ }\href@noop {}
  {\bibfield  {journal} {\bibinfo  {journal} {Soviet Physics JETP}\ }\textbf
  {\bibinfo {volume} {13}},\ \bibinfo {pages} {1120--1125} (\bibinfo {year}
  {1961})}\BibitemShut {NoStop}%
\bibitem [{\citenamefont {Bogle}\ and\ \citenamefont
  {Symmons}(1959)}]{Bogle1959}%
  \BibitemOpen
  \bibfield  {author} {\bibinfo {author} {\bibfnamefont {G.~S.}\ \bibnamefont
  {Bogle}}\ and\ \bibinfo {author} {\bibfnamefont {H.~F.}\ \bibnamefont
  {Symmons}},\ }\bibfield  {title} {\enquote {\bibinfo {title} {Paramagnetic
  resonance of {Fe$^{3+}$} in sapphire at low temperatures},}\ }\href@noop {}
  {\bibfield  {journal} {\bibinfo  {journal} {Proceedings of the Physical
  Society.}\ }\textbf {\bibinfo {volume} {73}},\ \bibinfo {pages} {531--532}
  (\bibinfo {year} {1959})}\BibitemShut {NoStop}%
\end{thebibliography}%

\end{document}